\documentstyle[12pt]{article}

\newcommand{\be}{\begin{eqnarray}}
\newcommand{\ee}{\end{eqnarray}}
\newcommand{\e}{\varepsilon}
\newcommand{\D}{\partial}

\begin{document}
\sloppy
\gdef\journal#1, #2, #3, 1#4#5#6{{#1~}{#2} (1#4#5#6) #3}
\gdef\ibid#1, #2, 1#3#4#5{{#1} (1#3#4#5) #2}

\begin{flushright}  
IPNO-TH 96-02\\
Oslo-SHS-96-8 \\
hep-th/9606080\\
\end{flushright}
\vskip 5mm


{\centerline{\large\bf Equation of State for Exclusion Statistics in a 
Harmonic Well }}
\vskip 1cm
{\centerline {\large Serguei B. Isakov$^{a,1,2}$ and  St\'ephane Ouvry$^b$}}
\vskip 3mm
{\centerline {\it ${}^{a }$Centre for Advanced Study, Norwegian Academy of
Science and Letters}}
{\centerline{\it Drammensveien 78, 0271 Oslo, Norway}}
{\centerline{\it ${}^b$Division de Physique Th\'eorique ${}^3$,
IPN, Orsay Fr-91406, France}}

\vskip 1cm

\centerline {\bf Abstract}
\smallskip

We consider the equations of state 
for systems of particles with exclusion statistics in a harmonic well.
Paradygmatic examples are
noninteracting particles obeying ideal fractional exclusion statistics placed
in 
 (i) a harmonic well on a line, and
(ii) a harmonic well in the Lowest Landau Level (LLL) of an exterior 
magnetic field. We show their identity  with (i) the Calogero model 
 and (ii) anyons in the LLL of an exterior 
magnetic field  and in
a harmonic well.

\bigskip
\noindent


\vfil
\noindent
$^{1}$Permanent address: Medical Radiological Research Center, Obninsk,
Kaluga Region 249020, Russia. \\
$^2$Address from  1 August 1996: Institute of Physics,  University of Oslo,
P.O. Box 1048 Blindern, N-0316 Oslo, Norway.\\
$^3$Unit\'e de recherche du CNRS associ\'ee aux Universit\'es Paris 11 et
Paris 6.\\
\newpage   
Haldane's definition of fractional statistics based on a generalized 
exclusion principle \cite{Hal1} when applied {\it locally} in phase space 
results in single state 
statistical distributions for {\it ideal} fractional exclusion statistics (IFES)
\cite{I1,Wu}. IFES first appeared in the litterature for anyons in the 
lowest Landau 
level (LLL) \cite{dVO-PRL94} of an exterior  magnetic field, i.e. for a gas
of particles with a degenerate one body spectrum. On the other hand, in one dimension, 
IFES can be modelled with  inverse square 
interactions \cite{ES-Cal-Suth}. 
The relevant systems are 
the Sutherland model (particles on a circle) \cite{Suth}  and the 
Calogero model (particles on a line in a harmonic potential) \cite{Cal}.


To derive 
the equation of state and the virial expansion of  
interacting systems modelling IFES \cite{dVO-PRL95,MS-PRL95},
one can start from a system placed in a harmonic potential
and then use a thermodynamical prescription to obtain the equation of state
of the original system in the infinite box. 
The harmonic potential is then referred to as a ``long distance regulator''.
This  procedure, originally proposed for anyons \cite{toto},  
has been put on general ground in \cite{Ol,dVO-PRL95}.
In particular, this is precisely 
in this way that the equation of state for anyons in the LLL was 
originally derived \cite{dVO-PRL94} 
(for  several species of particles see \cite{IMO}.) 
The same procedure can be  applied to 1d particles 
with an inverse square interactions:  virial
coefficients for this  model have been derived
\cite{dVO-PRL95} 
starting from the energy levels of the Calogero model in a harmonic well 
and using an adequate 1d thermodynamic limit
prescription. The latter result was also verified 
using the equivalence of the Sutherland model to a system of free 
particles obeying IFES \cite{IAMP}. 


In this note, we address the question of the equation of state for exclusion 
statistics in a harmonic potential, without considering the transition to a 
system in the infinite box limit. We refer to this case as a ``physical'' harmonic 
potential. What we have in mind are  mesoscopic quantum dots systems
where  finite volume effects are supposed to be described by a 
harmonic well.
Since there is no well defined volume in this system,
the equation of state (or the potential 
$\Omega$) depends on the particle number $N$, the harmonic potential 
frequency $\omega$, 
and other parameters as well. A similar form for 
the equation of state have already  been discussed 
 in the Thomas-Fermi approximation for 1d IFES particles in an external 
potential \cite{SB}.

We start by assuming a certain temperature scaling for  the one-particle partition
function of  non interacting
IFES particles. In this context, the equation of state is valid both for
particles in an infinite  box and in a physical harmonic well.
As examples, we discuss non  interacting particles in
(i) a 1d harmonic well and
(ii) in a 2d harmonic well in the LLL of an external magnetic field. Both these cases correspond to a constant density of
levels in energy. We compare these equations of state with those for
(i) the Calogero model and (ii) anyons in a harmonic well in the LLL of an
external magnetic field.
We  discuss more generally a specific form of the equation of state for IFES 
with a constant density of states for systems both in a box and  
in a harmonic well. 
Introducing   the effective volume occupied by the gas in a  
harmonic well at a given temperature, we  obtain finally an 
equation   
of state in a physical harmonic well quite similar to those in a box.


Ideal fractional exclusion statistics  can be  defined by the 
single-state grand partition function 
\be
[\xi(x_i)]^{g-1}[\xi(x_i)-1]=x_i  \;,
\label{xi}\ee
where $\varepsilon_i$ is the energy of the state $i$ and  
$x_i$ is the Gibbs factor $x_i\equiv e^{\beta(\mu-\e_i)}$,  
$\beta=1/{k_{\scriptscriptstyle{\rm B}}}T$.  

The statistics parameter $g=0$ corresponds to Bose and 
$g=1$ to Fermi statistics. The distribution function  
$n(x_i)=x_i(\partial/\partial x_i) \ln \xi(x_i)$ is 
connected with $\xi(x_i)$ by the bilinear relation 
\be
\frac{1}{n(x_i)}=\frac{1}{\xi(x_i)-1} + g  \;.
\label{bilinear}\ee
Expanding  in powers of $x_i$ yields 
\be
\ln \xi(x_i)=\sum_{k=1}^{\infty} \frac{Q_k}{k}x_i^k\;, \quad 
Q_k=\prod_{l=1}^{k-1} \left(1-g\,{k\over l}\right)=
\frac{\Gamma(k-gk)}{\Gamma(k)\Gamma(1-k)}\; .
\label{lnxi_i}\ee
By summing $\ln \xi(x_i)$ and $n(x_i)$ 
over $i$, one obtains the expansions  
\be
-\beta\Omega=\sum_{k=1}^{\infty} b_k z^k\; ,\quad N=\sum_{k=1}^{\infty} k b_k z^k 
\label{expansions}\ee
($z=e^{\beta\mu}$ is  the fugacity), with the cluster coefficients 
\be
\quad b_k=\frac{Q_k}{k}Z_1(k\beta )\;,  
\label{b_k}\ee
 $Z_1(\beta)=\sum_i e^{-\beta\e_i}$ is the one-particle partition function. 

Without specifying in detail the system at this stage, we 
assume that the partition function 
$Z_1(\beta)$ scales with the inverse temperature $\beta$ as
\be
Z_1(k\beta)\simeq\frac{e^{-k\beta\e_0}}{k^{1+\delta}}Z'_1(\beta) \;.
\label{Z1-k}\ee
This scaling is relevant for systems with a gap $\varepsilon_0$ in the single particle  energy spectrum, 
with  one-particle partition function 
$Z_1(\beta)=e^{-\beta\varepsilon_0}Z'_1(\beta)$. 
This factorization will indeed materialize in  the thermodynamic limit for various 
physical systems as we will see below. 

Taking into account (\ref{Z1-k}), the cluster coefficients 
(\ref{b_k}) become 
\be
b_k=\frac{Q_k}{k^{2+\delta}}e^{-k\beta\e_0}Z'_1(\beta) \;.
\label{b_k=}\ee 
One deduces from (\ref{expansions}) the ``virial expansion'' 
\be
-\beta\Omega=\sum_{k=1}^{\infty} A_k {{N^k}\over {[Z'_1(\beta)]^{k-1}}}\;, 
\label{`vir-exp'}\ee  
with the dimensionless ``virial coefficients''
\be
A_1=1,\quad
A_2=\frac1{2^{2+\delta }}(2g-1),\quad
A_3=[(4^{-1-\delta }-2\cdot3^{-2-\delta })+g(g-1)(4^{-\delta }-3^{-\delta })],
\dots 
\label{A}\ee

We now  specialize on the particular case $\delta =0$, which
can be completely analyzed. This case is of particular interest since it
implies a $1/\beta$ scaling for $Z_1(\beta)$, which means a constant density
of states in energy.
One can already see from (\ref{A}) that $\delta=0$ is special since it implies 
that
$A_3$ does not depend on the statistical parameter $g$.

Comparing the second expansion in (\ref{expansions}), 
with the cluster coefficients (\ref{b_k=}), with (\ref{lnxi_i}), one can write 
\be
 N=Z'_1(\beta)\, \ln \xi(z') \;,
\label{N=}\ee
where $z'=ze^{-\beta\e_0}$. 
Regarding Eq.~(\ref{N=}) as determining   $z'$ as a function of $N$, 
we obtain from (\ref{expansions})
\be
-{{\D \,\beta\Omega}\over {\D N}}=\frac{1}{z}{{\D z'(N) }\over {\D N}}N \;. 
\label{deriv}\ee   
 To calculate $\D z'(N) / \D N$ (or 
$\D N(z') / \D z'$), we use (\ref{N=}) and 
(\ref{xi})--(\ref{bilinear}). 
Noting that  $(1/N(z'))\D N(z') / \D z'=n(z')/z'$, 
we find an equation of state  
\be
-\frac{\D\,\beta \Omega}{\D N}=\frac{N}{Z'_1(\beta)} 
\left( \frac1{e^{N/Z'_1(\beta) }-1}+ g  \right) \;, 
\label{deriv=}\ee
and  upon   integration, finally,   
\be
-\beta\Omega= 
\int_0^{N}\frac{N'}{Z'_1(\beta)}
\frac{d N'}{(e^{N'/Z'_1(\beta) }-1)} 
+ \frac12 g  \frac{ N^2}{Z'_1(\beta)} \; . 
\label{eqst}\ee
Expanding this, we obtain the virial expansion   
\be
-\beta\Omega=N \left\{ 1+\frac14(2g-1)\frac{N}{Z'_1(\beta)}
+\sum_{k=2}^{\infty}\frac{{\cal B}_k}{(k+1)!}
\left(\frac{N}{Z'_1(\beta)} \right)^k
\right\} \;,
\label{virexp}\ee
where 
${\cal B}_k$ are the Bernoulli numbers  (${\cal B}_2=\frac16$, 
${\cal B}_4=-\frac{1}{30}$), vanishing for $k$ odd. 
It appears that the statistical parameter
$g$ only enters the equation of state via the second virial coefficient
(\cite{dd}).

Let us now examine how the equations of state (\ref{`vir-exp'}) and (\ref{eqst}) 
can be relevant to both systems  in a box  and in  a physical  harmonic well. 

We first consider a gas of 
free (spinless) particles, with generic dispersion law 
$\e(p)=\e_0+ap^{\sigma}$ occupying a box of volume $V$ in $d$ dimensions. 
In the thermodynamic limit the one-particle partition function is 
\be
Z_1(k\beta)=e^{-k\beta\e_0}
\frac{\Gamma(1+d/\sigma)V}
{ (2\sqrt{\pi}{})^d \Gamma(1+{\textstyle{1 \over 2}}d)
(ak\beta)^{-d/\sigma} }\; , 
\label{Z1}\ee
satisfying (\ref{Z1-k}), with $\delta=d/\sigma -1$.
In this case the virial expansion (\ref{`vir-exp'}) becomes the usual
virial expansion for a system in a box with  pressure  $P$ given by
\be
\beta P= \sum_{k=1}^{\infty} a_k \rho^k, \quad \rho=N/V\;, 
\label{vir-exp-usual}\ee
with the (dimensional) virial coefficients  
\be
a_k=A_k \left( {{V}\over {Z'_1(\beta)}} \right)^{k-1} ,
\label{a-dim}\ee
where $A_k$ are given in (\ref{A}).
Note that the virial coefficients (\ref{a-dim}) coincide with those obtained 
in \cite{IAMP} for $\e_0=0$ showing that the presence of a gap in 
the particle dispersion does not affect the equation of state.   

The constant density of states case,
$\delta=d/\sigma -1=0$, describes for example chiral particles on a line with linear dispersion with a gap, 
\be
\e=\e_0+vp\; , \quad p\geq 0 \; .
\label{e-chiral}\ee
The equation of state is (\ref{eqst}) (or (\ref{virexp})) with 
\be
Z'_1(\beta)={{V}\over {2\pi{} v \beta}}\; .
\label{Z1-1d}\ee      

Let us now turn to systems in a ``physical'' harmonic well. 
We restrict to  cases with a constant density of states, i.e. 
we  consider noninteracting  (nonrelativistic) particles 
that  occupy single particle levels 

\medskip
\begin{tabular}{rl}
(i) & in a 1d harmonic well 
 $V({x})=\frac12 m\omega^2 {x}^2$ ;\\

(ii)& in the LLL of an external  magnetic field $B$ 
       in a 2d harmonic well.  
\end{tabular}
\medskip

\noindent
In a quantum dot 
language, where $\omega$ is small (for example with respect to 
the cyclotron frequency), but non vanishing, the harmonic well 
is supposed to encode the finite size effects of the sample.

The one particle energy levels are
\be
\e_{\ell}=\e_0+\ell \varpi\;, \quad   \ell\geq 0\;, 
\label{1part-energy}\ee
where for the 1d harmonic model 
\be
\e_0=\frac12 \omega\;, \quad \varpi=\omega \;,
\label{Cal}\ee
whereas for  the 2d LLL harmonic model, one has
\be
\e_0=\sqrt{\omega_c^2+\omega^2}\;, \quad \varpi
=\sqrt{\omega_c^2+\omega^2}-\omega_c \;, 
\label{LLL}\ee
or, to leading order in $\omega^2/\omega_c^2$, 
\be
\e_0=\omega_c\;, \quad \varpi=\frac{\omega^2}{2\omega_c} \;.
\label{LLL-leading}\ee
It follows from (\ref{1part-energy}) that the one-particle partition function is 
\be
Z_1(k\beta)= 
\frac{e^{-k\beta\e_0}}{1-e^{-k\beta\varpi}}
=\frac{e^{-k\beta\e_0}}{k\beta\varpi}
\left(1+{\textstyle \frac12}(k\beta\varpi)+{\cal O}[(k\beta\varpi)^2] \right)\;.
\label{Z_1}\ee
Thus, to leading order in $k\beta\varpi$, 
the partition function (\ref{Z_1}) scales as in (\ref{Z1-k}), 
with $\delta=0$ and 
\be
Z'_1(\beta)=1/\beta\varpi \;.
\label{Z1'}\ee
 This leads to  
the equation of state (\ref{eqst}),(\ref{virexp}).  
If one considers the correction terms ${\cal O}(k\beta\varpi)$  
 in (\ref{Z_1}), they might lead to  corrections to the virial coefficients 
of very high order, $N$ and above. However, if the virial expansion converges, 
these corrections are negligible inside the radius of convergence.  

It is interesting to note that 
for 1d IFES particles  in a harmonic well,  the equation 
of state (\ref{eqst}),(\ref{virexp}) with (\ref{Z1'}) and (\ref{Cal})
coincides with that obtained in the Thomas-Fermi 
approximation \cite{SB}. We stress that the  
present derivation only uses the condition   
$\beta{}\omega\ll 1$  
when the discreteness of energy levels becomes  inessential.

Interacting systems modelling IFES are known to be 
the 1d Calogero model and 2d anyons in the LLL of an
external
magnetic field. An harmonic well is added to these systems to define a proper 
thermodynamic limit. It would be  interesting to investigate what happens if the
 harmonic well  becomes now physical.
The $N$-body energy 
spectrum is given by \cite{dVO-PRL95}
\be
\sum_{\ell}n_\ell\varepsilon_{\ell} +{1\over2}N(N-1)\varpi g \;,
\label{levels}\ee
where $n_\ell$ are nonnegative integers
with the constraint $\sum_{\ell}n_{\ell}=N$, and where
$\e_{\ell}$ is given by (\ref{1part-energy}).
For anyons, $g=\alpha\in [0,1]$ is the anyonic statistical parameter; for the
Calogero model, $g=\alpha \geq 0$   defines  the singular
two-body potential
 $\alpha(\alpha-1)/(x_i-x_j)^2$ and 
 specifies
 the coinciding point (short distance) behavior
of the $N$-body wave function,
$\Psi\propto |x_i-x_j|^\alpha$ as $|x_i-x_j|\to 0$.

The $N$-body partition function corresponding to (\ref{levels}) is 
\be
Z_N=e^{-{1\over2}\beta N(N-1)\varpi g} 
\prod_{n=1}^N{e^{-\beta\varepsilon_0}\over 1-e^{-n\beta\varpi}} \;.
\label{ZN}\ee
Remarkably,  the cluster coefficients 
obtained form (\ref{ZN}) are \cite{dVO-PRL94}
\be
\quad b_k=e^{-k\beta\e_0}\frac{Q_k}{k^2 \beta\varpi} 
(1+{\cal O}(\beta\varpi))\;,   
\label{b-interacting}\ee
having, to leading order in $\beta\varpi$, the  form (\ref{b_k=}) with 
$\delta=0$ and $Z'_1(\beta)$ given by (\ref{Z1'}). As above, 
the terms of order ${\cal O}(\beta\varpi)$
in (\ref{b-interacting})
are negligible inside the radius of convergence   of  the virial expansion.
It follows that both  equations of state of the Calogero and     
 LLL-anyon model in a 2d harmonic well are given by 
(\ref{eqst}),(\ref{virexp}) with $Z'_1(\beta)$  determined by (\ref{Z1}), i.e. they are
identical to the equation of states 
for noninteracting particles in a 1d harmonic well or 
in a 2d LLL harmonic well.

This happens to be general feature of equations of states 
in a physical harmonic well for various physical systems 
modelling  IFES with a constant density of levels:  
only the second virial coefficient  depends on the statistics parameter, in
agreement with the general statements  for 
free IFES particles in a box \cite{IAMP}. 
  
As a specific example, consider 
the equation of state for a  model of a chiral field  
on a circle  proposed in \cite{HLV}. This model was constructed by  
mapping the second quantized  LLL anyon model on a circle.
The energy levels of this model are given by  (\ref{levels})
with the identification
\be
\frac{\omega^2}{2\omega_c}=\frac{2\pi}{V}v \;, 
\label{ident}\ee
where $V$  is the length of the circle. If the  harmonic  
potential is assumed to materialize in the anyon droplet of radius $V/2\pi$ 
by the action of an electric field, 
the velocity $v$ can then be   interpreted as the drift velocity $E/B$
on the edge 
 (the velocity of the edge excitations), where  the 
electric field on the edge is  $E=(m/e)\omega^2 R$. 
The thermodynamic limit is understood as $\omega\to 0$, $V\to\infty$ whereas 
$v$ is kept fixed. The spectrum (\ref{levels}), altogether  with the identification 
(\ref{ident}), yield the cluster coefficients (\ref{b_k=}) with 
$\delta=0$, $\e_0=\omega_c$ ($Z'_1(\beta)$ is given by (\ref{Z1-1d})). 
It follows that the equation of state
is (\ref{eqst}),(\ref{virexp}), that is those of free chiral 1d IFES particles 
with  dispersion (\ref{e-chiral}).   
This conclusion is consistent with those of \cite{IV},
namely the model of a chiral 
field on a circle 
\cite{HLV} admits an interpretation in terms of IFES.

We finally note that (\ref{eqst}) (or (\ref{virexp})) 
for a physical harmonic well, with $Z'_1(\beta)$ given by  
(\ref{Z1'}),   
can be viewed as an equation of state
connecting the average pressure to the average particle density
in the harmonic 
well. The local pressure $P(\bf x)$  in  
a slowly varying  $d$-dimensional harmonic potential is defined as 
\be
\beta\Omega=\int d^d{\bf x}\beta P({\bf x})\;.
\label{Xi}\ee
The average pressure is then
\be
\beta \langle P\rangle \equiv 
\left(\frac{{}\beta\omega}{\lambda_T}\right)^d
\int d^d{\bf x} \beta P({\bf x})\;,
\label{Pressure}\ee
where $\lambda_T={} \sqrt{2\pi\beta/m}$ is the thermal de Broglie wavelength.  
The local particle density 
$\rho({\bf x})$ being normalized as
$\int d^D{\bf x} \rho({\bf x})=N$, one can define the 
average density as 
\be
\langle \rho\rangle \equiv 
\left(\frac{{}\beta\omega}{\lambda_T}\right)^d
\int d^d{\bf x} 
 \rho({\bf x})\;.
\label{Density}\ee
Introducing the  effective volume occupied by the gas at temperature $T$
by 
\be
V_{\rm eff}= \left(\frac{\lambda_T}{{}\beta\omega}\right)^d, 
\label{V_eff}\ee
we have $V_{\rm eff}\sim R_{\rm eff}^d$, where 
$R_{\rm eff}$ is determined by 
\be
\frac12 m\omega^2 R_{\rm eff}^2\simeq T \;. 
\label{R_eff}\ee 
It follows that (\ref{virexp}) can then be rewritten as 
a virial expansion in a box of volume $V_{\rm eff}$
(cf (\ref{vir-exp-usual}), (\ref{a-dim})): 
 \be
\beta \langle P\rangle  =\langle \rho\rangle \left(1 
 +\frac14 (2g-1){\beta\varpi V_{\rm eff}}{\langle \rho\rangle}+
\sum_{k=2}^{\infty} \frac{{\cal B}_k}{(k+1)!}{(\beta\varpi V_{\rm eff})^k}
{\langle  \rho\rangle}^{k} \right) \,.
\label{vir-exp}\ee
For noninteracting particles in a 1d harmonic well and for the Calogero 
model, $\beta\varpi V_{\rm eff}=\lambda_T$, whereas for   
noninteracting LLL particles in a 2d harmonic well and  for 
LLL anyons in  a 2d harmonic well,  $\beta\varpi V_{\rm eff}=2\pi/eB$. 
For the Calogero model, the equation of state (\ref{vir-exp}) was 
conjectured in \cite{dVO-PRL95}.
Note that for the LLL anyon model, the condition $1 \ll 2\beta \omega_c$ guarantees that the particles, 
when confined in the quantum dot of size $R_{\rm eff}$ given by (\ref{R_eff}),  
do not reach the second Landau level.  

\smallskip
{\bf Acknowledgements:} 
We would like to thank J.M. Leinaas and S. Mashkevich for 
valuable discussions.





\begin{thebibliography}{40}
\bibitem{dd} for discussions of systems where only the second virial coefficient
is affected by the statistics, 
see A. Dasni\`eres de Veigy, Nucl. Phys. B458[FS], 533,  1996; see also Ref. [9]
\end{thebibliography}
\end{document}